\documentclass[aps,prd,reprint,groupedaddress,nofootinbib,preprintnumbers]{revtex4-2}

\usepackage{graphicx}
\usepackage{mathtools}
\usepackage{physics}
\usepackage{hyperref} 

\usepackage[disable]{todonotes}

\makeatletter
\g@addto@macro\bfseries{\boldmath}
\makeatother

\newcommand{\m}[1]{\mathrm{#1}}

\let\phi\varphi
\let\epsilon\varepsilon

\begin{document}

\preprint{[HU-EP-25/35-RTG, DESY-25-140]}

\title{Position-space sampling for local multiquark operators in lattice QCD using distillation and the importance of tetraquark operators for $T_{cc}(3875)^+$}


\author{Andres Stump}
\email{andres.stump@hu-berlin.de}
\affiliation{Institut für Physik, Humboldt-Universität zu Berlin,\\
Zum Großen Windkanal 2, 12489 Berlin, Germany}

\author{Jeremy~R.~Green}
\email{jeremy.green@desy.de}
\affiliation{Deutsches Elektronen-Synchrotron DESY,\\
  Platanenallee 6, 15738 Zeuthen, Germany}


\date{\today}

\begin{abstract}
  Obtaining hadronic two-point functions is a central step in spectroscopy calculations in lattice QCD. This requires solving the Dirac equation repeatedly, which is computationally demanding. The distillation method addresses this difficulty by using the lowest eigenvectors of the spatial Laplacian to construct a subspace in which the Dirac operator can be fully inverted. This approach is efficient for nonlocal operators such as meson-meson and baryon-baryon operators. However, local multiquark operators with four or more (anti)quarks are computationally expensive in this framework: the cost of contractions scales with a high power of the number of Laplacian eigenvectors. We present a position-space sampling method within distillation that reduces this cost scaling by performing the momentum projection only over sparse grids rather than the full spatial lattice. We demonstrate the efficiency of this unbiased estimator for single-meson, single-baryon and local tetraquark operators. Using Wilson-clover fermions at the $SU(3)$-flavor-symmetric point, we apply this method to study the importance of local tetraquark operators for extracting the finite-volume $T_{cc}(3875)^+$ spectrum. To this end, we extend a large basis of bilocal $DD^*$ and $D^*D^*$ scattering operators by including local tetraquark operators. The inclusion of local operators leads to significant shifts in the estimates of several energy levels. Finally, we show the effect of these shifts on the $DD^*$ scattering phase shift from a single-channel $s$-wave Lüscher analysis.
\end{abstract}

\keywords{[Keywords?]}

\maketitle

\section{Introduction}

In recent decades, numerous exotic hadrons have been discovered, including the $\chi_{c1}(3872)$ and other charmonium-like states, as well as various tetraquarks and pentaquarks~\cite{Husken:2024rdk, Chen:2022asf}. A prominent example is the $T_{cc}(3875)^+$ tetraquark, which was observed and studied at LHCb~\cite{LHCb:2021vvq, LHCb:2021auc}. It was found in the $D^0D^0\pi^+$ mass spectrum just below the $D^{*+}D^0$ threshold and is an $I(J^P)=0(1^+)$ state. With a minimal quark content of $cc\overline{u}\overline{d}$, it does not fit into the quark model picture and is therefore clearly an exotic hadron. Another interesting exotic hadron is the $d^*(2380)$ dibaryon discovered by the WASA-at-COSY collaboration~\cite{PhysRevLett.106.242302}. It is an excited state of the deuteron with $I(J^P) = 0(3^+)$, lying about 80~MeV below the $\Delta\Delta$ threshold.

These discoveries call for theoretical predictions of the masses, widths, and other properties of such exotic hadrons. Lattice quantum chromodynamics (QCD) is the only known method to provide such predictions from first principles. Since in lattice QCD one always works in a finite box, it is necessary to relate finite-volume quantities obtained there to infinite-volume observables. For single-particle states that are stable within QCD, finite volume effects on the energy are exponentially suppressed by the box size times the smallest mass scale, i.e.\ the pion mass~\cite{Luscher:1985dn}. However, most hadrons are not QCD stable and can decay into other hadrons. In this case, the finite-volume contributions only decay with inverse powers of the box size and have to be taken into account. One way of dealing with these finite-volume effects is to use Lüscher's finite-volume quantization conditions~\cite{Luscher:1986pf,Luscher:1990ux}. They relate the low-lying finite-volume energies to the scattering amplitude in infinite volume. The positions of the poles in this amplitude are then used to extract the masses and widths of the hadrons via amplitude analysis.

The current best strategy for computing the low-lying finite-volume spectrum is the variational method~\cite{Blossier:2009kd}. It uses a basis of interpolating operators, all carrying the quantum numbers of the desired hadron, to construct a correlator matrix. By solving a generalized eigenvalue problem (GEVP), the lowest few energy levels can be extracted at large Euclidean times. To suppress the influence of excited states, one wants to use many such operators, and since the inner structure of exotic hadrons is generally not known, they should have different spin, color and spatial structures to capture the low-lying spectrum well. Therefore, we want to use nonlocal scattering operators to describe states that resemble weakly bound ``molecular'' states, as well as local operators for more deeply bound states. Not including such a variety of operators can lead to an inaccurate finite-volume spectrum~\cite{Wilson:2015dqa}.

An efficient framework for computing correlation functions of nonlocal operators is distillation~\cite{HadronSpectrum:2009krc}. It employs Laplacian Heaviside (LapH) smearing, in which the lowest eigenvectors of the spatial Laplacian define a subspace that allows to fully invert the Dirac operator. Distillation has been used successfully for two and three meson scattering~\cite{Wilson:2015dqa, Moir:2016srx, Briceno:2017qmb, Woss:2019hse, Lang:2022elg, Boyle:2024hvv, Yan:2024gwp, Mohler:2013rwa, Fischer:2020yvw, Hansen:2020otl, Fischer:2020jzp, Dawid:2025doq}, tetraquarks~\cite{Padmanath:2022cvl, Whyte:2024ihh, Prelovsek:2025vbr, Shrimal:2025wbu}, meson-baryon scattering~\cite{Lang:2012db, Hackl:2024whw, Bulava:2022vpq, BaryonScatteringBaSc:2023ori, BaryonScatteringBaSc:2023zvt} and baryon-baryon scattering~\cite{Horz:2020zvv, PhysRevLett.127.242003, Xing:2025uai, BaSc:2025yhy}. On the other hand, two-point functions of local multiquark operators with four or more (anti)quarks can be obtained more easily using standard approaches such as the point-to-all method. Nevertheless, they have also been computed within the distillation framework~\cite{Cheung:2017tnt, Ortiz-Pacheco:2023ble,Prelovsek:2025vbr}. It is challenging to use these local operators in distillation because one needs to construct high-rank tensors, which make the contractions computationally expensive. A possible solution to this problem was recently proposed~\cite{Lang:2024syy}.

In this paper, we present a different approach for reducing the cost of local operators in distillation. This method uses position-space sampling, and we applied it to the quantum numbers relevant for the $T_{cc}$ tetraquark.

This paper is organized as follows: In Sec.~\ref{sec:distillation_method}, we give a brief overview of the distillation method and show why local multiquark operators are expensive in this framework. In Sec.~\ref{sec:position_space_sampling}, we present a position-space sampling method within distillation and show how it can be used to reduce the cost of local multiquark operators. In Sec.~\ref{sec:numerical_investigation}, we investigate the efficiency of this method for single-meson, single-baryon and local tetraquark operators, and analyze the importance of the latter for the $T_{cc}$ tetraquark. We end with our conclusion in Sec.~\ref{sec:conclusion}.

\section{Distillation Method} \label{sec:distillation_method}

In hadron spectroscopy, field smearing is crucial for reducing the coupling of hadron operators to high-energy states. For the quark fields, we can use a smearing kernel $K(t)$ that acts on spin, color and position space but does not mix different time slices. The smeared quark field is then given by
\begin{equation}
  \psi_\m{sm}(t) = K(t)\,\psi(t).
\end{equation}

A widely used quark smearing method is LapH smearing, employed in distillation~\cite{HadronSpectrum:2009krc,Morningstar:2011ka}. To establish notation and make this paper self-contained, we briefly review LapH smearing and the distillation method. The smearing kernel $K$ in LapH smearing is constructed from the lowest eigenvectors of the (stout-smeared~\cite{Morningstar:2003gk}) gauge-covariant Laplacian $\Delta(t)$. More precisely, we set $K = VV^\dagger$ where $V$ contains the lowest $N_v$ eigenvectors of $\Delta(t)$. Since all eigenvalues $\lambda^{(k)}(t)$ of the Laplacian are negative, we order them as
\begin{equation}
  -\lambda^{(1)}(t) \leq -\lambda^{(2)}(t) \leq -\lambda^{(3)}(t) \leq \dots,
\end{equation}
and define $V$ as
\begin{equation}
  V(t) = \left(v^{(1)}(t), v^{(2)}(t), \dots, v^{(N_v)}(t)\right),
\end{equation}
where $v^{(k)}(t)$ is the eigenvector corresponding to the eigenvalue $\lambda^{(k)}(t)$. This is a $3|\Lambda_3| \times N_v$ matrix, where $\Lambda_3$ is the spatial lattice and $3$ is the number of colors. The Laplacian acts trivially in spin space, so $V$ and the smearing kernel $K$ constructed from it do as well. Since the Laplacian is defined in terms of the gauge links, its eigenvectors, and thus $V(t)$, are time dependent. We call the Laplacian eigenvectors Laplace modes and the index $k$ the Laplace mode index.

Using only the $N_v$ lowest Laplace modes is similar to imposing a cutoff $\sigma^2$ on its eigenvalues. This way, the smearing kernel can be written in compact form as
\begin{equation}
  K = \Theta(\sigma^2 + \Delta)
\end{equation}
(cf.~\cite{Morningstar:2011ka}), using the Heaviside function $\Theta$. The difference from the former definition is that the number of eigenvalues of $-\Delta(t)$ below $\sigma^2$ does not need to be constant across different time slices. However, in~\cite{Morningstar:2011ka} it was shown that this number changes very little. Therefore, in numerical simulations we use a constant number of vectors to define $K$.

The Laplacian is covariant under gauge transformations, rotations on the lattice, parity and charge conjugation, and since the smearing kernel $K$ is defined in terms of its eigenvectors, it preserves these symmetries as well. Thus, no relevant symmetry of the quark fields is broken by LapH smearing.

When we smear the quark fields with $K$, we get the smeared quark propagator (with flavor $f$) as
\begin{align}
\begin{split}
  S_{f, \,\m{sm}}(t', t) &= K(t') \cdot S_f(t', t) \cdot K(t)^\dagger \\
    &= V(t')V(t')^\dag \cdot S_f(t', t) \cdot V(t) V(t)^\dag
\end{split}
\end{align}
in terms of the unsmeared propagator $S_f = D_f^{-1}$. Here, $D_f$ is the Dirac operator for the quark with flavor $f$. This is a $12|\Lambda_3| \times 12|\Lambda_3|$ matrix, and it is not feasible to invert it fully. However, $K$ is a projector onto the space spanned by the $N_v$ lowest Laplace modes. This so-called LapH subspace is $4N_v$ dimensional, and $N_v$ is typically a few tens or a few hundreds; consequently, the Dirac operator can be fully inverted in this subspace. This is done by introducing the so-called perambulator, defined as 
\begin{equation}
    (\tau_f)_{\alpha\beta}(t', t) = V(t')^\dag \cdot (S_f)_{\alpha\beta}(t', t) \cdot V(t).
\end{equation}
Here we have explicitly written out the spin indices. The perambulator is a $4N_v \times 4N_v$ matrix, and it can be computed by solving the Dirac equation for each Laplace mode $v^{(k)}(t)$ and spin index $\beta$ at the source, and then projecting the result to the Laplace modes at the sink.

Finally, the smeared propagator is given by
\begin{equation} \label{eq:smeared_propagator}
  \big(S_{f, \,\m{sm}}\big)_{\alpha\beta}(t', t) = V(t') \cdot (\tau_f)_{\alpha\beta}(t', t) \cdot V(t)^\dag,
\end{equation}
in terms of the perambulator.

As an example, we consider the charged pion to illustrate how distillation works in practice. For this, we use the operator
\begin{equation} \label{eq:pion_operator}
  \mathcal{O}^\pi(t) = \sum_{\vb*{x} \in \Lambda_3} e^{-i\vb*{p}\cdot\vb*{x}} (\overline{d} \gamma_5 u)(\vb*{x}, t),
\end{equation}
which is projected to momentum $\vb*{p}$.\footnote{We suppress the dependence on the momentum $\vb*{p}$ in most cases.} Its unsmeared two-point function is given by
\begin{align} \label{eq:pion_two_point_function}
\begin{split}
  C(t', t)
    =& \sum_{\vb*{x}', \vb*{x} \in \Lambda_3} e^{-i\vb*{p}\cdot(\vb*{x}'-\vb*{x})} \\
    &\hspace{15pt} \cdot \ev{\tr\left[
        S_u(\vb*{x}', t'; \vb*{x}, t)\cdot
        S_d(\vb*{x}', t'; \vb*{x}, t)^\dag
  \right]}_G,
\end{split}
\end{align}
where $\ev{\dots}_G$ is the expectation value over the gauge fields only. To get the smeared two-point function, we replace the propagators by the smeared ones according to Eq.~\eqref{eq:smeared_propagator}. This gives us
\begin{equation} \label{eq:smeared_pion_two_point_function}
    C_\m{sm}(t', t) = \ev{\tr\left[ 
            \Phi(t')\cdot \tau_u(t', t)\cdot
            \Phi(t)^\dag\cdot \tau_d(t', t)^\dag
        \right]}_G,
\end{equation}
where we have collected the momentum projection and the Laplace modes in the \textit{mode doublets} $\Phi(t)$ defined as
\begin{equation} \label{eq:mode_doublet}
  \Phi^{(k, l)}(t) = \sum_{\vb*{x} \in \Lambda_3} e^{-i\vb*{p}\cdot\vb*{x}} \, v_a^{(k)}(\vb*{x}, t)^*  \,v_a^{(l)}(\vb*{x}, t).
\end{equation}
They form an $N_v\times N_v$ matrix. Here we have explicitly written out the Laplace mode indices $(k, l)$ and the color index $a$ in addition to the position index $\vb*{x}$. Using these mode doublets in numerical simulations is easier than recomputing or storing the Laplace modes. We see that distillation has turned the pion two-point function into simple matrix-matrix multiplications in the LapH subspace.

Before moving on to more complicated operators, we address the question of how many Laplace modes to use. As mentioned above, we can relate the number of Laplace modes $N_v$ to a cutoff $\sigma^2$ on the eigenvalues of the Laplacian. This cutoff corresponds to a certain smearing radius, which we can tune to obtain a fast convergence of the effective energy for a given two-point function. Once we have chosen a certain $N_v$, we want to keep the physical smearing radius constant as we change the lattice spacing $a$ and the volume $V$. In~\cite{Morningstar:2011ka}, it was shown that to keep the physical smearing radius constant, we have to keep the number of Laplace modes $N_v$ proportional to the physical volume $V = a^3|\Lambda_3|$. Thus, going to larger volumes increases the computational cost for computing the perambulators. In addition, the contractions become more expensive. For simple meson two-point functions, this is manageable, but for local multiquark operators with four or more (anti)quarks, it becomes problematic.

\subsection{Local multiquark operators in distillation} \label{sec:local_multiquark_operators_in_distill}

As for mesons, we can use distillation for baryons~\cite{HadronSpectrum:2009krc, Morningstar:2011ka, PhysRevLett.127.242003} by following the same steps as before: we first compute an unsmeared baryon two-point function and replace the propagators with smeared ones. Then we want to somehow absorb the Laplace modes and the momentum projection into a tensor. For mesons, we ended up with the mode doublets, which form a rank-2 tensor. But for baryons, we get a rank-3 tensor with one Laplace mode index for each quark. These \textit{mode triplets} are defined as
\begin{equation}
  \Psi^{(k, l, m)}(t) = \sum_{\vb*{x} \in \Lambda_3} e^{-i\vb*{p}\cdot\vb*{x}} \, 
      \epsilon_{abc} \, v_a^{(k)}(\vb*{x}, t) v_b^{(l)}(\vb*{x}, t) v_c^{(m)}(\vb*{x}, t).
\end{equation}
Similarly, we can proceed and compute two-point functions of bilocal meson-meson and baryon-baryon scattering operators. We can reuse the mode doublets and triplets we have already defined for them, since the mesons and baryons are each projected to a separate momentum.

\begin{figure}
  \includegraphics[width=0.45\textwidth]{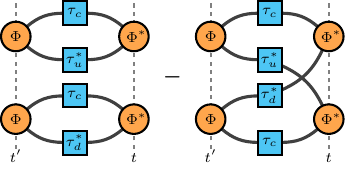}
  \includegraphics[width=0.195\textwidth]{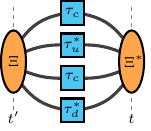}
  \caption{\label{fig:corr_in_distill}Tensor network diagrams of two-point functions in distillation of operators relevant for the $T_{cc}$ tetraquark; a bilocal $DD^*$ (top) and a local tetraquark operator (bottom). For the bilocal $DD^*$ operator, the two terms from the Wick contraction are topologically distinct, whereas for the local tetraquark operator, they have the same topology. The spin structure is suppressed.}
\end{figure}

The computational cost of computing two-point functions of meson and bilocal meson-meson operators scales as $N_v^3$ since the contractions consist of matrix-matrix multiplications in the LapH subspace. An example for such a meson-meson operator is the bilocal $DD^*$ operator that is relevant for the $T_{cc}$ tetraquark. We visualized the tensor network diagram of its two-point function in the top panel of Fig.~\ref{fig:corr_in_distill}.
One can see that it consists only of traces in the LapH subspace.

For baryon and bilocal baryon-baryon operators, the cost for computing two-point functions asymptotically scales as $N_v^4$ as we have to contract matrices (the perambulators) with rank-3 tensors (the mode triplets). 

\begin{figure}
  \includegraphics[width=0.5\textwidth]{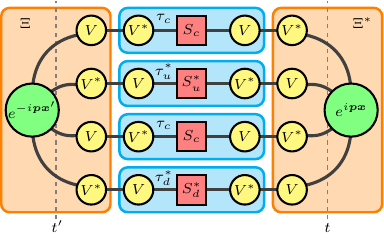}
  \includegraphics[width=0.5\textwidth]{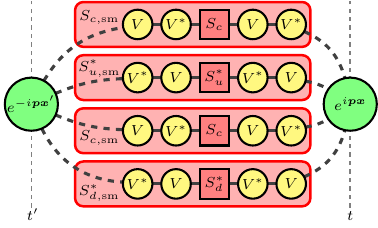}
  \caption{\label{fig:corr_in_distill_and_pos_space_samp}Tensor network diagrams of the same tetraquark two-point function as in the bottom panel of Fig.~\ref{fig:corr_in_distill}. Here, all tensors involved in distillation are displayed to visualize the order in which they are contracted. In the conventional distillation method (top), the perambulators $\tau_f$ (with flavor $f$) are constructed by contracting the propagators $S_f$ with the neighboring $V$ matrices. The remaining $V$'s are contracted with $e^{-i\vb*{p}\vb*{x}'}$ which gives the rank-4 tensor $\Xi$. In the position-space sampling method (bottom), the propagators are contracted with all four $V$'s connected to it, giving the smeared propagators $S_{f,\m{sm}}$. The contraction with the exponentials is then performed only over subspaces of the spatial lattice (indicated by the dashed lines). The spin and color structure is suppressed in these diagrams.}
\end{figure}

However, as mentioned in the introduction, we are also interested in local multiquark operators. For example, for the $T_{cc}$ tetraquark, a generic local tetraquark operator has the form
\begin{equation}
  \mathcal{O}^T(t) \sim \sum_{\vb*{x} \in \Lambda_3}
  e^{-i\vb*{p}\cdot\vb*{x}} (cc\bar{u}\bar{d})(\vb*{x}, t),
\end{equation}
where we have suppressed the spin and color structure. It could, for example, be a local $DD^*$ or a diquark-antidiquark operator. When computing two-point functions, we again collect the momentum projection and the Laplace modes into a tensor, which in this case has rank~4. For a local $DD^*$ operator, it is given by
\begin{equation}
\begin{split}
	\Xi^{(k,l,m,o)}(t) =& \\
		\sum_{\vb*{x} \in \Lambda_3} e^{-i\vb*{p}\cdot\vb*{x}}
		& v_a^{(k)}(\vb*{x}, t)^* v_a^{(l)}(\vb*{x}, t) \; 
    v_{b}^{(m)}(\vb*{x}, t)^* v_{b}^{(o)}(\vb*{x}, t),
\end{split}
\end{equation}
whereas for a diquark-antidiquark operator the color structure differs.
The tensor network diagram of the corresponding two-point function in distillation is displayed in the bottom panel of Fig.~\ref{fig:corr_in_distill}. In Fig.~\ref{fig:corr_in_distill_and_pos_space_samp}, we visualize the order in which the propagators and the Laplace modes are contracted to first obtain the perambulators and $\Xi$.
Then these tensors are contracted to arrive at the two-point function.

Consequently, the overall computational cost for this two-point function scales as $N_v^5$. This is very expensive, especially for large physical volumes. An additional problem is the memory required to store the rank-4 tensor $\Xi$.

The problem becomes more severe when we want to use local hexaquark operators for dibaryons such as the $d^*(2380)$. These operators are of the form
\begin{equation}
  \mathcal{O}^H(t) \sim \sum_{\vb*{x} \in \Lambda_3} 
        e^{-i\vb*{p}\cdot\vb*{x}} (uuuddd)(\vb*{x}, t),
\end{equation}
where again we have suppressed the spin and color structure. It is not clear whether these operators are relevant for the low-lying finite-volume energies of a certain dibaryon system. Previous lattice studies found that they are not important for the low-lying nucleon-nucleon energies~\cite{Amarasinghe:2021lqa, Detmold:2024iwz, BaSc:2025yhy}, but it is possible that they are relevant for other dibaryons such as the $d^*$ and the $H$~\cite{Francis:2018qch}.

To use these local hexaquark operators in distillation, we would have to construct a rank-6 tensor, and the computational cost would scale as $N_v^7$. This is prohibitively expensive even for a reasonably small number of Laplace modes such as $N_v=32$.

The reason why the computational cost for these local multiquark operators scales so badly is the order in which the tensors are contracted. By summing over the position indices first, we construct high-rank tensors that make the contractions over the Laplace mode indices expensive. To avoid these high-rank tensors, we can change the contraction order as it is visualized in the bottom panel of Fig.~\ref{fig:corr_in_distill_and_pos_space_samp} for a local tetraquark operator. This means we explicitly construct the smeared propagators as defined in Eq.~\eqref{eq:smeared_propagator} by summing over all Laplace mode indices first. The remaining contraction then involves a double sum over the position indices $\vb*{x}'$ and $\vb*{x}$ as in Eq.~\eqref{eq:pion_two_point_function}.

Let us analyze the cost scaling for this case. The computation of the smeared propagators from perambulators consists of two matrix-matrix multiplications with costs $N_v^2|\Lambda_3|$ and $N_v |\Lambda_3|^2$, and the double sum over the position space scales as $|\Lambda_3|^2$. As mentioned before, to have a constant smearing radius, we have to keep $N_v$ proportional to the physical volume $V = a^3|\Lambda_3|$. Consequently, for a constant lattice spacing, the asymptotic cost scaling is proportional to $N_v^3$, and it comes from computing the smeared propagators. This is much better than the $N_v^5$ and $N_v^7$ scaling for local tetraquark and hexaquark operators when constructing the aforementioned high-rank tensors, and it is independent of the number of quarks. But since $|\Lambda_3| \gg N_v$, the computation is still very expensive. Moreover, one would have to write the full Laplace modes to disk or recompute them for each contraction batch job; both options are impractical.

The solution we propose in this paper for these local multiquark operators is to use a position-space sampling method in addition to distillation. This reduces the cost of computing the smeared propagators and of the contractions.

\section{Position-space sampling} \label{sec:position_space_sampling}

The basic idea of the position-space sampling method is to compute the smeared propagators only on subspaces $\tilde{\Lambda}_3 \subset \Lambda_3$ rather than on the full spatial lattice $\Lambda_3$. More precisely, we choose two subspaces, $\tilde{\Lambda}_3'$ and $\tilde{\Lambda}_3$: the first for the sink and the second for the source. Then, we compute the smeared propagators $D_{f,\,\m{sm}}^{-1}(\vb*{x}',t'; \vb*{x},t)$ for all $\vb*{x}' \in \tilde{\Lambda}_3'$ and $\vb*{x} \in \tilde{\Lambda}_3$ from the perambulators and the Laplace modes, and contract them to obtain the two-point function. The subspaces $\tilde{\Lambda}_3'$ and $\tilde{\Lambda}_3$ have to be chosen with a certain scheme for each gauge configuration and source time $t$ if multiple sources are used. Before we look at possible sparsening schemes, we first illustrate how this position-space sampling method works.

Applying this sparsening approach to the charged pion Eq.~\eqref{eq:pion_two_point_function}, we obtain the smeared pion two-point function within position-space sampling $\tilde{C}_\m{sm}(t', t)$ as
\begin{widetext}
\begin{equation} \label{eq:pos_space_sampling_pion}
  \tilde{C}_\m{sm}(t', t) = \Big<
    \frac{|\Lambda_3|^2}{|\tilde{\Lambda}'_3||\tilde{\Lambda}_3|}\,
    \sum_{
      \mathclap{\substack{\vb*{x}' \in \tilde{\Lambda}'_3 \\
                \vb*{x} \in \tilde{\Lambda}_3}}
    }
    e^{-i\vb*{p}\cdot(\vb*{x}' - \vb*{x})} 
      \tr\left[
          S_{u,\,\m{sm}}(\vb*{x}',t'; \vb*{x},t) \cdot
          S_{d,\,\m{sm}}(\vb*{x}',t'; \vb*{x}, t)^\dag
      \right]\Big>_{G, \tilde{\Lambda}_3^{(\prime)}},
\end{equation}
\end{widetext}
where we have added the prefactor ${|\Lambda_3|^2/(|\tilde{\Lambda}'_3||\tilde{\Lambda}_3|)}$ to ensure the correct normalization.  The expectation value $\ev{\dots}_{G, \tilde{\Lambda}_3^{(\prime)}}$ is now an expectation value both over the gauge fields and over the two subspaces, since they do not have to be the same for all gauge configurations.

For general local multiquark operators, the procedure is the same, only the contraction of the smeared propagators is more complicated. In the tensor network diagram in the bottom panel of Fig.~\ref{fig:corr_in_distill_and_pos_space_samp}, we use dashed lines to indicate the contractions that are performed using position-space sampling.

To use this approach, we need to store the Laplace modes on the two subspaces $\tilde{\Lambda}_3'$ and $\tilde{\Lambda}_3$ for each gauge configuration. Various approaches for defining these subspaces have been used in the literature ~\cite{Detmold:2019fbk, Li:2020hbj}, though not combined with distillation. They can be grouped into two categories: sparse grids and random subspaces. The former uses some types of sparse grids, such as regular grids, where only every fourth or every eighth point in each spatial direction is used. With this sparsening approach, a relatively large separation between the points can be used, without introducing significant additional noise from field sparsening. However, the disadvantage is that it leads to an incomplete momentum projection, so the resulting estimator is biased (cf.~\cite{Li:2020hbj}).

The second approach uses fully random subspaces, meaning that the points are chosen randomly from the full spatial lattice $\Lambda_3$. The advantage of this approach is that it leads to a complete momentum projection, so the resulting estimator is unbiased. However, numerical investigations show that it leads to a larger variance compared to the sparse-grids method~\cite{Li:2020hbj}.

In this work, we propose to use a regular sparse grid that is randomly shifted. More precisely, we choose a point separation $N_\m{sep}$ that divides the spatial extent $N_s$ of the lattice and define the subspace
\begin{align} \label{eq:sparse_grid}
\begin{split}
  \tilde{\Lambda}_{3, \tilde{\vb*{x}}} = \{a \vb*{n} + \vb*{\tilde{x}}\, |&
    n_k = 0,\, N_\m{sep},\, 2N_\m{sep},\, \dots,\, N_s-N_\m{sep} \\
    & (k = 1, 2, 3) \},
\end{split}
\end{align}
with a random offset $\vb*{\tilde{x}} \in \Lambda_3$. The number of points in this subspace is ${|\tilde{\Lambda}_{3, \tilde{\vb*{x}}}| = |\Lambda_3|/{N_\m{sep}}^3}$. To properly sample the full lattice, we use a different random offset $\vb*{\tilde{x}}$ for each gauge configuration and source time $t$ for the sparse grid at the source (denoted by $\tilde{\Lambda}_{3, \tilde{\vb*{x}}}$). For the sparse grid at the sink (denoted by $\tilde{\Lambda}_{3, \tilde{\vb*{x}}'}$), we sample a single random offset $\vb*{\tilde{x}}'$ for each gauge configuration, and use it for all sink times $t'$. This allows us to preserve correlations between different sink times $t'$, yielding more precise effective energies.

The important difference from the sparse-grids method without a random offset is that this sparsening approach is unbiased. To show this, we first note that all two-point functions of local operators with definite momentum can be written as
\begin{equation}
  C(t', t) = \sum_{\mathclap{\vb*{x}', \vb*{x} \in \Lambda_3}}
    \ev{f(\vb*{x}',t'; \vb*{x},t)}_G.
\end{equation}
The function $f$ encompasses the (smeared) quark propagators and the exponential from the momentum projection. When applying our position-space sampling method, this expression becomes
\begin{equation} \label{eq:pos_space_sampling_general}
  \tilde{C}(t', t) = 
    \Big<
    \frac{|\Lambda_3|^2}{|\tilde{\Lambda}_{3, \vb*{0}}|^2}
    \;\sum_{
      \mathclap{\substack{\vb*{x}' \in \tilde{\Lambda}_{3, \tilde{\vb*{x}}'} \\
                \vb*{x} \in \tilde{\Lambda}_{3, \tilde{\vb*{x}}}}}
    }
    f(\vb*{x}',t'; \vb*{x},t)\Big>_{G, \tilde{\vb*{x}}^{(\prime)}}.
\end{equation}
Here, we have changed the notation for the expectation value to $\ev{\dots}_{G, \tilde{\vb*{x}}^{(\prime)}}$, since the expectation value over the sparse grids is just an average over all possible offsets $\vb*{\tilde{x}}'$ and $\vb*{\tilde{x}}$. Consequently, we have
\begin{equation}
  \ev{\dots}_{G, \tilde{\vb*{x}}^{(\prime)}}
    = \frac{1}{\abs{\Lambda_3}^2} \sum_{\tilde{\vb*{x}}', \tilde{\vb*{x}} \in \Lambda_3} \ev{\dots}_G.
\end{equation}

To make the dependence on $\vb*{\tilde{x}}'$ and $\vb*{\tilde{x}}$ in the sparse grids explicit, we can write
\begin{equation}
  \tilde{\Lambda}_{3, \tilde{\vb*{x}}^{(\prime)}} =
    \tilde{\Lambda}_{3, \vb*{0}} + \vb*{\tilde{x}}^{(\prime)}
\end{equation}
where the addition is understood elementwise. That way we can move the offsets $\vb*{\tilde{x}}'$ and $\vb*{\tilde{x}}$ from the sum into $f$. Using these properties, Eq.~\eqref{eq:pos_space_sampling_general} becomes 
\begin{equation}
  \tilde{C}(t', t) = 
    \frac{1}{|\tilde{\Lambda}_{3, \vb*{0}}|^2}
    \hspace{10pt}\sum_{\mathclap{\substack{\vb*{x}', \vb*{x} \in \tilde{\Lambda}_{3,\vb*{0}} \\
                                 \tilde{\vb*{x}}', \tilde{\vb*{x}} \in \Lambda_3}}}
    \ev{f(\vb*{x}' + \tilde{\vb*{x}}', t'; \vb*{x} + \tilde{\vb*{x}}, t)}_G.
\end{equation}
Due to the periodicity of the spatial lattice, we can perform the substitutions
\begin{equation}
  \tilde{\vb*{x}}' \rightarrow \tilde{\vb*{x}}' - \vb*{x}', \quad
  \tilde{\vb*{x}} \rightarrow \tilde{\vb*{x}} - \vb*{x}
\end{equation}
in the offsets. Then, the sum over $\vb*{x}'$ and $\vb*{x}$ becomes trivial, yielding a factor $|\tilde{\Lambda}_{3, \vb*{0}}|^2$. Finally, we obtain
\begin{equation}
  \tilde{C}(t', t) =
    \sum_{\mathclap{\tilde{\vb*{x}}', \tilde{\vb*{x}} \in \Lambda_3}}
    \ev{f(\tilde{\vb*{x}}', t'; \tilde{\vb*{x}}, t)}_G
    = C(t', t),
\end{equation}
i.e.\ we recover the sum over the full spatial lattice. Consequently, the position-space sampling estimator is unbiased.

We can also adapt this derivation to mixed two-point functions of a local and a bilocal operator. In that case, we only use position-space sampling for the local operator, i.e.\ only at the sink or the source. Hence, also for these mixed two-point functions, the position-space sampling estimator is unbiased.\footnote{In this case, the random offset in the sparse grid is not required to obtain an unbiased estimator.}

The point separation $N_\m{sep}$ in the sparse grids gives us control over the variance, where for $N_\m{sep} = 1$ we get the result from distillation without position-space sampling. Ideally, we want to choose $N_\m{sep}$ such that the uncertainty from position-space sampling is much smaller than the Monte Carlo error. We expect that the noise from position-space sampling depends primarily on the physical distance $aN_\m{sep}$ between the points as the lattice spacing and volume are changed. Thus, once we have chosen an $N_\m{sep}$ for a given operator, we want to keep $aN_\m{sep}$ constant as we change the lattice spacing and volume. This means we keep the number of points in the sparse grids $|\Lambda_3|/{N_\m{sep}}^3$ proportional to the physical volume $a^3|\Lambda_3|$, i.e.\ proportional to the number of Laplace modes $N_v$. As a result, we have the same $V^3$ cost scaling as for $N_\m{sep}=1$ (cf. Sec.~\ref{sec:local_multiquark_operators_in_distill}), but with a much smaller prefactor that depends on the chosen $N_\m{sep}$. In addition, the scaling toward the continuum ($a\rightarrow0$) is improved when keeping $aN_\m{sep}$ constant, compared to $N_\m{sep}=1$. Consequently, position-space sampling makes local multiquark operators with four or more (anti)quarks feasible within distillation, especially in large physical volumes.

\section{Numerical investigation} \label{sec:numerical_investigation}

In this Section, we show spectroscopy results from applying our position-space sampling method. First, we study the efficiency of this method and investigate which point separation $N_\m{sep}$ should be used for computing two-point functions of single-meson, single-baryon and local tetraquark operators. Then, we analyze the importance of the latter for the $T_{cc}$ tetraquark by comparing estimates of the low-lying finite-volume spectrum obtained with and without including local tetraquark operators. Finally, we investigate the effect of including these operators on the extracted $s$-wave $DD^*$ scattering phase shift, determined using Lüscher's finite-volume quantization conditions. All errors only include statistical uncertainties computed using the $\Gamma$ method~\cite{WOLFF2004143}.

\subsection{Lattice setup} \label{sec:lattice_setup}

The simulations were performed using two CLS~\cite{Bruno:2014jqa} gauge ensembles with $O(a)$-improved Wilson fermions at the $SU(3)$-flavor-symmetric point. This means the quark mass is approximately set to the average of the physical $u$, $d$ and $s$ quarks. The relevant parameters for the simulations are summarized in Table~\ref{tab:ensembles}. The resulting pion mass is $m_\pi \approx 420$~MeV on both ensembles. For the valence charm quark, the same fermion action was used. Its quark mass was tuned such that the $D$ meson mass approximately matches the average of the physical $D^0$, $D^+$ and $D_s^+$ masses, which is 1901~MeV. The unphysical pion mass results in a $D^*$ meson that is stable within QCD; this avoids three-particle decays of the $T_{cc}$. For the B450 ensemble, we also used a heavier-than-physical charm quark with hopping parameter $\kappa_c=0.115336$, resulting in a $D$ mass of $m_D = 2753$~MeV.
We used stout smearing~\cite{Morningstar:2003gk} in the spatial Laplacian.

\begin{table*}
  \caption{\label{tab:ensembles}Parameters of the gauge ensembles used for the simulations. The lattice spacing was determined in~\cite{Strassberger:2021tsu}. $N_{\text{cfg}}$ and $N_{\text{src}}$ are the number of gauge field configurations and sources per gauge configuration used. The point separation $N_\m{sep}$ denotes the value that was used for the $T_{cc}$ simulations.}
  \begin{ruledtabular}
  \begin{tabular}{lccccccccc}
    Ensemble & $N_s^3 \times N_t$ & $\beta$ & $a$ (fm) & $\kappa_u=\kappa_d=\kappa_s$ & $\kappa_c$ & $N_{\text{cfg}}$ & $N_{\text{src}}$ & $N_v$ & $N_\m{sep}$ \\
    \hline
    B450 & $32^3\times64$  & 3.46 & 0.0749 & 0.136890 & 0.126243 & 1612 & 8 & 32 & 8 \\
    N202 & $48^3\times128$ & 3.55 & 0.0633 & 0.137000 & 0.128423 &  899 & 8 & 68 & 8 \\
  \end{tabular}
  \end{ruledtabular}
\end{table*}

\subsection{$N_\m{sep}$ dependence of position-space sampling method}

To study the efficiency of our position-space sampling method, we computed two-point functions of single-meson, single-baryon and local tetraquark operators using different point separations $N_\m{sep}$ in the sparse grids. Then we investigated the change in the variance of both the effective energy $E_\m{eff}$ and the energy from a plateau fit to $E_\m{eff}$. When decreasing the point separation $N_\m{sep}$, we expect the variance to decrease until the error is dominated by the Monte Carlo error. From there on, decreasing $N_\m{sep}$ will not reduce the overall error further. That $N_\m{sep}$ value is the ideal point separation for numerical simulations, although one can choose a bigger $N_\m{sep}$ to make simulations affordable at the cost of a larger error. The goal of this investigation is to find that ideal $N_\m{sep}$.

We did this $N_\m{sep}$-dependence analysis mostly using the B450 ensemble (see Table~\ref{tab:ensembles}) to keep the computational cost lower. The sparse grids that we use for position-space sampling [cf. Eq.~\eqref{eq:sparse_grid}] require a point separation $N_\m{sep}$ that divides the spatial extent $N_s$ of the lattice. For B450 this value is $N_s = 32$, so the possible $N_\m{sep}$ are $N_\m{sep} = 1, 2, 4, 8, 16, 32$ where $N_\m{sep}=1$ means using the full lattice and $N_\m{sep}=32$ means using only one point from the spatial lattice on a given time slice. The maximum point separation $N_\m{sep}=32$ led to large noise such that the correlator data was not usable. Therefore, $N_\m{sep}=16$ is the biggest separation that we used.

\subsubsection{Single-meson and single-baryon operators}

\begin{figure*}
  \includegraphics[width=0.5\textwidth]{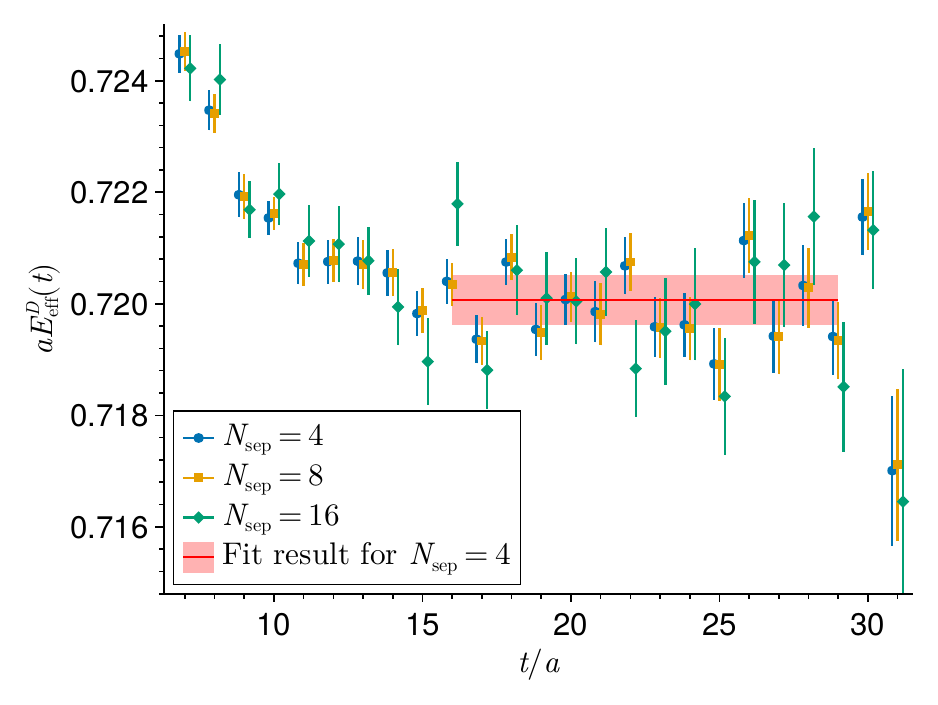}%
  \includegraphics[width=0.5\textwidth]{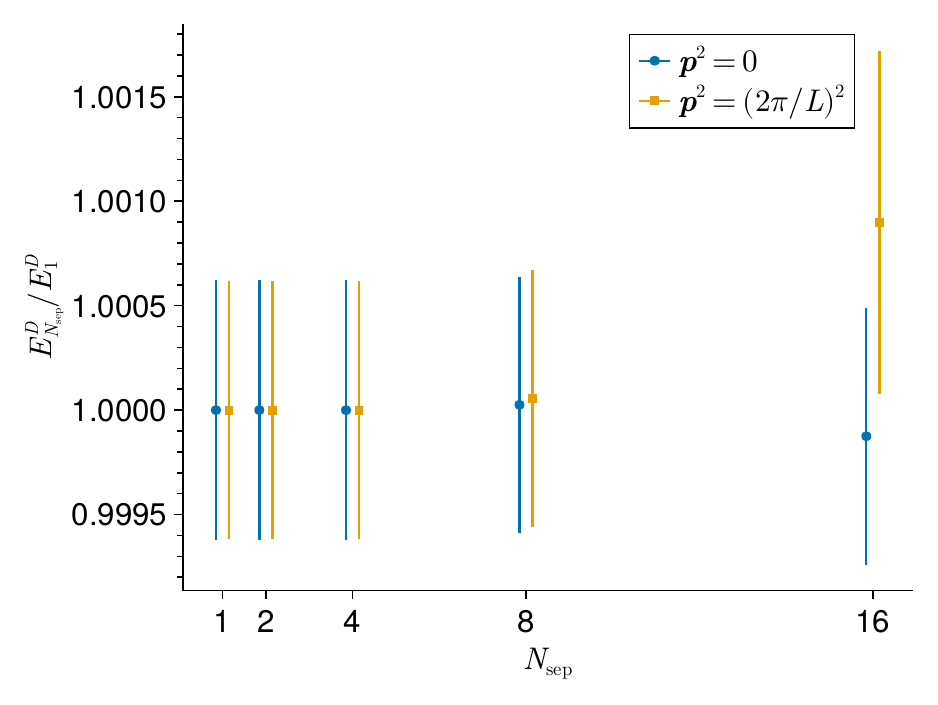}
  
  \includegraphics[width=0.5\textwidth]{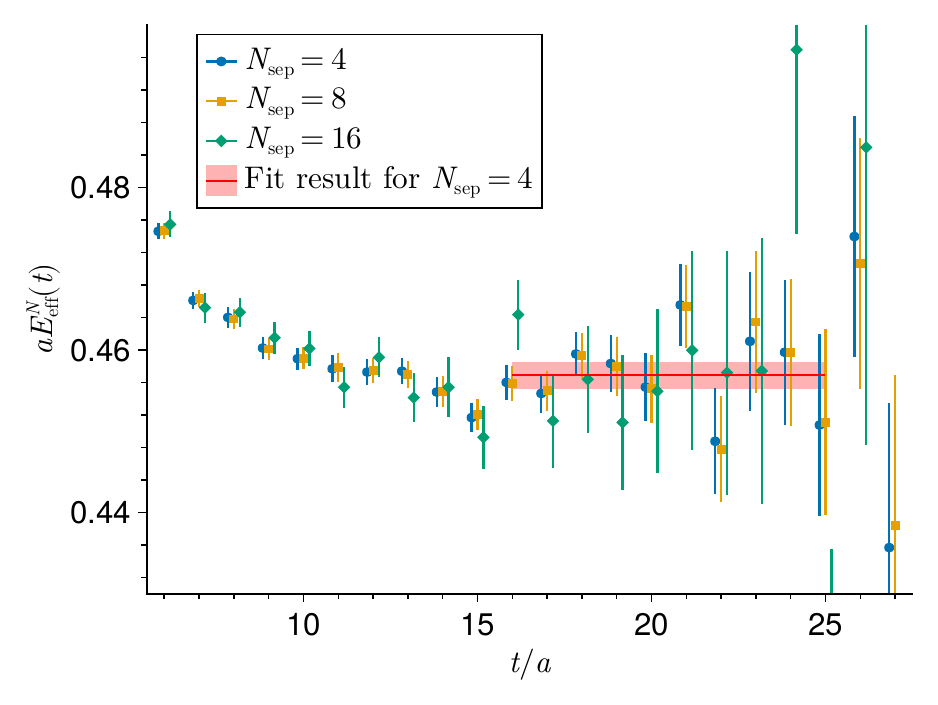}%
  \includegraphics[width=0.5\textwidth]{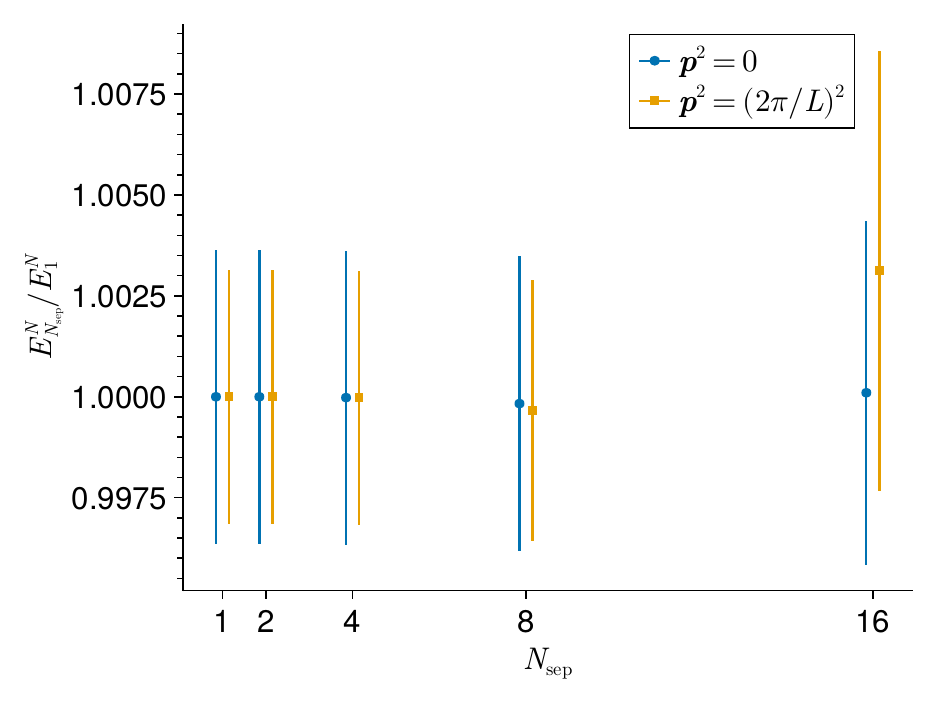}
  \caption{\label{fig:Nsep_analyis_for_D_and_N}Effective energies $E_\m{eff}^h$ for the $D$ meson ($h = D$, top left) and the nucleon ($h = N$, bottom left) at zero momentum and for different point separations $N_\m{sep}$. The red bands are results from plateau fits for $N_\m{sep}=4$. The right panel displays the energies $E_{N_\m{sep}}^h$ obtained from plateau fits to $E_\m{eff}^h$ for the $D$ meson (top right) and the nucleon (bottom right) as a function of $N_\m{sep}$. The energies $E_{N_\m{sep}}^h$ were computed using momenta with $\vb*{p}^2 = 0$ and $\vb*{p}^2 = (2\pi/L)^2$, and are normalized with the $N_\m{sep}=1$ value $E_1^h$.}
\end{figure*}

As an example single-meson system, we investigated the $D$ meson. For that, we used the pseudoscalar operator
\begin{equation} \label{eq:D_operator}
  \mathcal{O}^D(t) = \sum_{\vb*{x} \in \Lambda_3} e^{-i\vb*{p}\cdot\vb*{x}} (\overline{u} \gamma_5 c)(\vb*{x}, t)
\end{equation}
for the momenta $\vb*{p}=\vb*{0}$ and $\vb*{p} = \frac{2\pi}{L}(1, 0, 0)$, where $L = aN_s$ is the physical lattice size. The top left panel in Fig.~\ref{fig:Nsep_analyis_for_D_and_N} shows the effective energy $E_\m{eff}^D$ obtained from $D$ meson two-point functions with $\vb*{p}=\vb*{0}$ for $N_\m{sep}=4, 8, 16$.
For single mesons, we used the $\cosh$ definition of the effective energy. We see a significant decrease in error when going from $N_\m{sep}=16$ to $N_\m{sep}=8$. But the error does not decrease further when going to $N_\m{sep}=4$. This indicates that the error from position-space sampling is dominated by the Monte Carlo error already for $N_\m{sep}=8$.

The top right panel in Fig.~\ref{fig:Nsep_analyis_for_D_and_N} shows the $D$ meson energies $E_{N_\m{sep}}^D$ obtained from plateau fits to $E_\m{eff}^D$ as a function of $N_\m{sep}$. We used the same fit range for all $N_\m{sep}$ to isolate the $N_\m{sep}$ dependence, and we normalized the result with the $N_\m{sep}=1$ value $E_1^D$. Unlike for the effective energy, the zero-momentum plateau fit shows no significant change in error up to $N_\m{sep}=16$; this indicates that the decreased errors in the effective energy are associated with increased correlations between different $t$. For $\vb*{p}^2 = (2\pi/L)^2$, the error in the energy $E_{N_\m{sep}}^D$ is constant for $N_\m{sep} \leq 8$ and increases for $N_\m{sep}=16$. For both momenta, all $E_{N_\m{sep}}^D$ values are consistent with the $N_\m{sep}=1$ value, which is expected since the position-space sampling estimator is unbiased.

We repeated this analysis for the nucleon $N$ using the operator
\begin{equation}
  \mathcal{O}^N(t)
		= \sum_{\vb*{x} \in \Lambda_3} e^{-i\vb*{p}\cdot\vb*{x}}
			\left[
				\epsilon_{abc} \; P_+ u_a 
				\left(u_b^T \, C\gamma_5P_+ \, d_c \right)
			\right](\vb*{x}, t),
\end{equation}
where $C$ is the charge conjugation matrix, $P_+$ is the positive parity projector, and the Roman letters are color indices. Here we also used momenta with $\vb*{p}^2 = 0$ and $\vb*{p}^2 = (2\pi/L)^2$, but for the latter, we averaged over the six spatial directions to improve the signal. The obtained results are shown in the bottom panels of Fig.~\ref{fig:Nsep_analyis_for_D_and_N}. As for the $D$ meson, we see a clear decrease in the error of the zero-momentum effective energy when going from $N_\m{sep}=16$ to $N_\m{sep}=8$. Decreasing the point separation any further does not lead to a reduction in the error. Similarly, the error of the energies $E_{N_\m{sep}}^N$ obtained from plateau fits to $E_\m{eff}^N$ are stable for $N_\m{sep} \leq 8$. For $N_\m{sep} = 16$, the zero-momentum energy shows a slight error increase, whereas the $\vb*{p}^2 = (2\pi/L)^2$ value increases significantly. Nonetheless, the change in error between $N_\m{sep}=16$ and $N_\m{sep}=8$ in the effective energy is much bigger than in the fit result, indicating again the increased correlation in the data. The same analysis for the pion and the $D^*$ meson leads to similar results.

We conclude that position-space sampling works well for two-point functions of single-meson and -baryon operators, i.e.\ we can use a large point separation $N_\m{sep}$ in the sparse grids without observing an increase in the error of observables. Consequently, on the B450 ensemble and for these observables, the ideal point separation is $N_\m{sep} = 8$.

\subsubsection{Tetraquark operators}

Position-space sampling starts to yield large cost savings for local multiquark operators with four or more (anti)quarks. Therefore, we now investigate its $N_\m{sep}$ dependence for local tetraquark operators relevant for the $T_{cc}$ tetraquark. It is a $I(J^P) = 0(1^+)$ state with minimal quark content $cc\bar{u}\bar{d}$. Simple local tetraquark operators that carry these quantum numbers are the local $DD^*$ operator
\begin{equation} \label{eq:operator_local_DDstar}
  T_i^{DD^*}(t) = \sum_{\vb*{x} \in \Lambda_3} e^{-i\vb*{p}\cdot\vb*{x}}
      (\overline{u} \gamma_5 c \; \overline{d} \gamma_i c)(\vb*{x}, t)
      - \{u \leftrightarrow d\},
\end{equation}
the local $D^*D^*$ operator
\begin{equation} \label{eq:operator_local_DstarDstar}
  T_i^{D^*D^*}(t) = \sum_{\vb*{x} \in \Lambda_3} e^{-i\vb*{p}\cdot\vb*{x}}
      \epsilon_{ijk}(\overline{u} \gamma_j c \; \overline{d} \gamma_k c)(\vb*{x}, t)
\end{equation}
and the local diquark-antidiquark operator
\begin{equation} \label{eq:operator_local_diq}
  T_i^\m{diq}(t) = 
		\sum_{\vb*{x} \in \Lambda_3} e^{-i\vb*{p}\cdot\vb*{x}}
		(\epsilon_{abc} \, c_b^T \, C\gamma_i \, c_c \;
		\epsilon_{ade} \,\overline{u}_d \, C\gamma_5 \, \overline{d}_e^T)(\vb*{x}, t).
\end{equation}
Such local tetraquark operators have already been incorporated in simulations~\cite{Cheung:2017tnt, Junnarkar:2018twb,Prelovsek:2025vbr}. The diquark-antidiquark operator is in the $(\vb*{\overline{3}}_c \otimes \vb*{3}_c)_{\vb*{1}_c}$ color representation, but there is also a diquark-antidiquark operator in the $(\vb*{6}_c \otimes \vb*{\overline{6}}_c)_{\vb*{1}_c}$ representation. The latter is usually discarded in favour of the former because of the repulsive interactions in the diquark $\vb*{6}_c$ and the antidiquark $\vb*{\overline{6}}_c$~\cite{Huang:2023jec}. Using Fierz transformations, the diquark-antidiquark operator $T_i^\m{diq}$
can be represented as a linear combination of products of two color singlet currents of the form
\begin{equation} \label{eq:fierz_two_color_singlet_currents}
  \sum_{\vb*{x} \in \Lambda_3} e^{-i\vb*{p}\cdot\vb*{x}}
      (\overline{u} \Gamma_A c \; \overline{d} \Gamma_B c)(\vb*{x}, t)
\end{equation}
for various combinations of elements $\Gamma_A$, $\Gamma_B$ of the Clifford algebra~\cite{Nieves:2003in, Padmanath:2015era}. Hence, this operator is related to the local $DD^*$ and the local $D^*D^*$ operator. However, since we do not use a complete basis of operators of the type given in Eq.~\eqref{eq:fierz_two_color_singlet_currents}, the diquark-antidiquark operator is still linearly independent.

In this work, we only consider the zero-momentum case, i.e.\ $\vb*{p} = \vb*{0}$, therefore the operators $T_i^{DD^*}$, $T_i^{D^*D^*}$ and $T_i^\m{diq}$ belong to the $T_1^+$ irreducible representation (irrep) of the octahedral group $O_h$. For the $N_\m{sep}$-dependence analysis, we only consider the local $DD^*$ and the local diquark-antidiquark operator. Due to their similar structure, we expect the local $D^*D^*$ operator to behave similarly to the local $DD^*$ operator.

These tetraquark operators are not optimized, and the energy levels for this multiparticle system are closely spaced. Therefore, we do not expect the excited states to decay before the signal is lost, implying that they do not correspond to genuine finite-volume energies. To still extract observables which we can use to compare different $N_\m{sep}$ values, we performed plateau fits to the effective energies at late times. Again, we used the same plateau range across all $N_\m{sep}$.

\begin{figure}
  \includegraphics[width=0.5\textwidth]{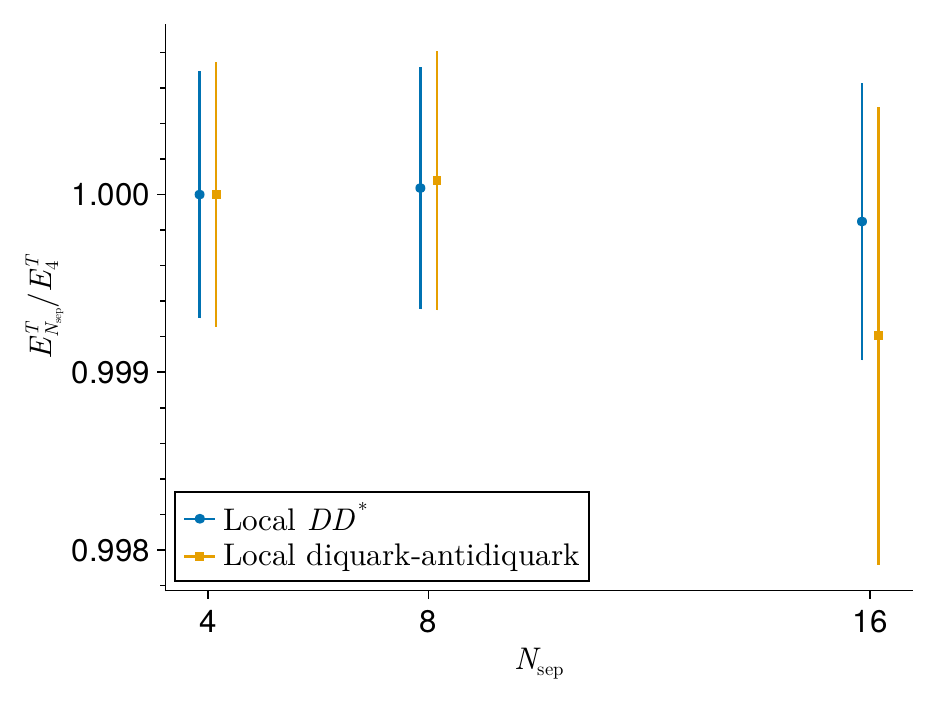}
  \caption{\label{fig:Nsep_analyis_for_Tcc}Plateau values $E_{N_\m{sep}}^T$ from plateau fits to the effective energy of the local $DD^*$ and diquark-antidiquark two-point functions vs. the point separation $N_\m{sep}$. The plateau values are normalized with the $N_\m{sep}=4$ value $E_4^T$.}
\end{figure}

The resulting plateau values $E_{N_\m{sep}}^T$ as a function of the point separation $N_\m{sep}$ are displayed in Fig.~\ref{fig:Nsep_analyis_for_Tcc}.
The cost for computing these two-point functions scales as ${N_\m{sep}}^{-6}$, and they are more expensive than the single-hadron two-point functions. Therefore, we only used the point separations $N_\m{sep} = 4, 8, 16$, and we used the best estimate $E_4^T$ to normalize the results. As for the nucleon, there is a decrease in error when going from $N_\m{sep} = 16$ to $N_\m{sep} = 8$, but then the error stabilizes. The effective energy shows the same behavior, although the reduction in error from $N_\m{sep} = 16$ to $N_\m{sep} = 8$ is more significant there. Hence, also for local tetraquark operators, we have a significant increase in correlation between different $t$ when reducing the point separation.

We conclude that also for local tetraquark operators the position-space sampling method works well and that the optimal point separation for these operators is $N_\m{sep} = 8$ on B450.

The other ensemble that we used for the $T_{cc}$ simulations is the N202 (see Table~\ref{tab:ensembles}). It has the lattice size $N_s = 48$, so possible and reasonable point separations are $N_\m{sep} = 4, 6, 8, 12, 16$. Its lattice spacing is $\sim15$\% smaller than that of B450. This means if we want to keep $aN_\m{sep}$ constant, we would have to use a point separation $N_\m{sep} \sim 9.5$. Consequently, we should use either $N_\m{sep} = 8$ or $N_\m{sep} = 12$. Our numerical investigation for $N_\m{sep} = 8, 12$ and 16 showed that the plateau values $E_{N_\m{sep}}^T$ for the local tetraquark operators stabilize already at $N_\m{sep} = 12$. But there is still a small error reduction in the effective energy when going from $N_\m{sep} = 12$ to $N_\m{sep} = 8$. Therefore, we also choose $N_\m{sep} = 8$ for the N202 ensemble. The results for the $T_{cc}$ presented in the next section were all obtained using this point separation.

\subsection{Results for $T_{cc}$}

In this Section, we present our spectroscopy results relevant for the $T_{cc}$. We begin by analyzing the importance of local tetraquark operators for extracting the finite-volume spectrum, then we show their effect on the resulting $DD^*$ $s$-wave scattering phase shift.

\subsubsection{Importance of local operators for finite-volume spectrum}

The importance of local tetraquark operators for extracting the finite-volume $cc\overline{u}\overline{d}$, isospin-0, $T_1^+$ spectrum has been discussed in the literature~\cite{Ortiz-Pacheco:2023ble,Prelovsek:2025vbr}. The authors find that these operators, more precisely the diquark-antidiquark operator, have little effect on the ground-state energy estimate but a significant impact on that of the first excited-state energy. In these two works, the rank-4 tensor $\Xi$ mentioned in Sec.~\ref{sec:local_multiquark_operators_in_distill} was used to compute two-point functions of local tetraquark operators within distillation. Because of their cost scaling with $N_v^5$, a smaller number of eigenvectors was used for the local tetraquark operators than for the bilocal meson-meson operators. This is different in our simulations, where we used the same $N_v$ for all operators.

For our importance analysis, we used a large basis of bilocal $DD^*$ and $D^*D^*$ operators in addition to the three local tetraquark operators defined in Eqs.~(\ref{eq:operator_local_DDstar})--(\ref{eq:operator_local_diq}). The bilocal $DD^*$ operators are constructed from linear combinations of operators of the form
\begin{align} \label{eq:operator_bilocal_DDstar}
\begin{split}
  \mathcal{O}_{i, \vb*{p}}^{DD^*}(t) = &
		\sum_{\vb*{x}_1, \vb*{x}_2 \in \Lambda_3}
		\hspace{-0.2cm} e^{-i\vb*{p}\cdot(\vb*{x}_1 - \vb*{x}_2)}
		(\overline{u} \gamma_5 c)(\vb*{x}_1, t) \; (\overline{d} \gamma_i c)(\vb*{x}_2, t) \\
    &- \{u \leftrightarrow d\},
\end{split}
\end{align}
and the bilocal $D^*D^*$ operator from
\begin{align} \label{eq:operator_bilocal_DstarDstar}
\begin{split}
  \mathcal{O}_{i, \vb*{p}}^{D^*\!D^*}(t) =& \\
		\sum_{\vb*{x}_1, \vb*{x}_2 \in \Lambda_3}
		\hspace{-0.2cm} &e^{-i\vb*{p}\cdot(\vb*{x}_1 - \vb*{x}_2)}
		\epsilon_{ijk}(\overline{u} \gamma_j c)(\vb*{x}_1, t) \; (\overline{d} \gamma_k c)(\vb*{x}_2, t).
\end{split}
\end{align}
Both were combined with the appropriate momentum structures to end up in the $T_1^+$ irrep (all of them are in the rest frame). The guiding principle for constructing these bilocal scattering operators was the following: we included as many operators as there are (degenerate) noninteracting energy levels on a given momentum shell.

To systematically investigate the effect of the different operator types and momentum shells on the spectrum, we organized the $DD^*$ and the $D^*D^*$ operators on a given momentum shell in groups that we call
\begin{equation} \label{eq:operator_group_notation}
  D^{(*)}(\vb*{k}^2)D^*(\vb*{k}^2)\;\{N_\m{ops}\},
\end{equation}
where $\vb*{k} = \frac{L}{2\pi}\vb*{p}$ is the integer momentum and $N_\m{ops}$ is the number of operators in this group. For the $DD^*$ type operators we used the momentum shells $\vb*{k}^2 = 0, 1, 2$ with $N_\m{ops} = 1, 2, 3$ operators, respectively. For the $D^*D^*$ we used $\vb*{k}^2 = 0, 1$ with $N_\m{ops} = 1, 2$, respectively. In total, those are nine bilocal meson-meson operators. Together with the three local tetraquark operators, whose group we call $T\;\{3\}$, the basis consists of 12 operators. For our investigation, we ordered these groups as follows: first the local tetraquark operators $T\;\{3\}$, then the $D^{(*)}D^*$ operator groups ordered by increasing noninteracting $D^{(*)}D^*$ energy. In this order, we extended the basis one operator group at a time, and in each step we extracted the low-lying energies. Then we repeated the same analysis without the tetraquark operators. By comparing the different energy estimates, we can deduce how important a certain operator group is for obtaining the correct spectrum.

To extract the low-lying energy levels, we used the variational method outlined in~\cite{Blossier:2009kd}. This means we solved the GEVP
\begin{equation} \label{eq:GEVP}
  C(t) v_n(t, t_0) = \lambda_n(t, t_0)\; C(t_0) v_n(t, t_0)
\end{equation}
for the correlator matrix
\begin{equation} \label{eq:correlator_matrix}
	C_{ij}(t) = \ev{\mathcal{O}_i(t) \mathcal{O}_j(0)^\dagger},
\end{equation}
where $\mathcal{O}_i$ are the  operators in the given basis. From the eigenvalues $\lambda_n(t, t_0)$ we computed the effective energy
\begin{equation} \label{eq:effective_energy}
	E_{\m{eff},n}(t) = -\frac{1}{a}\log(\frac{\lambda_n(t+a, t_0)}{\lambda_n(t, t_0)})
\end{equation}
by setting $t_0/a = \lceil t/(2a)\rceil$ where $\lceil\cdot\rceil$ is the ceiling function. That way, the condition $t_0 \geq t/2$ holds, which ensures a suppression of the excited states with $O\big(e^{-(E_{N_\m{ops}}-E_n)t}\big)$ in $E_{\m{eff},n}(t)$~\cite{Blossier:2009kd}. The energy levels $E_n$ could, in principle, be extracted directly from $E_{\m{eff},n}(t)$. However, the $E_n$ are strongly correlated with the noninteracting $DD^*$ energies. To reduce statistical errors, we therefore extracted the energies using plateau fits to the effective energy differences
\begin{equation}
  \Delta E_{\m{eff},n}(t) = E_{\m{eff},n}(t) - m^D_\m{eff}(t) - m^{D^*}_\m{eff}(t),
\end{equation}
where $m^D_\m{eff}(t)$ and $m^{D^*}_\m{eff}(t)$ are the effective masses of the $D$ and $D^*$ meson, respectively. This gives us the energy differences $\Delta E_n = E_n - m_D - m_{D^*}$ where $m_D$ and $m_{D^*}$ are the $D$ and $D^*$ masses. We computed these masses separately from plateau fits to $m^D_\m{eff}$ and $m^{D^*}_\m{eff}$ to obtain the $E_n$.

This method is very similar to performing fits to ratios of correlators, which can lead to fake plateaus~\cite{Iritani:2016jie}. To avoid this, we performed fits in a region where both the two-particle effective energy and the single-particle effective masses have reached a plateau. This approach was advocated in~\cite{Yamazaki:2017jfh}.

\begin{figure*}
  \includegraphics[width=0.5\textwidth]{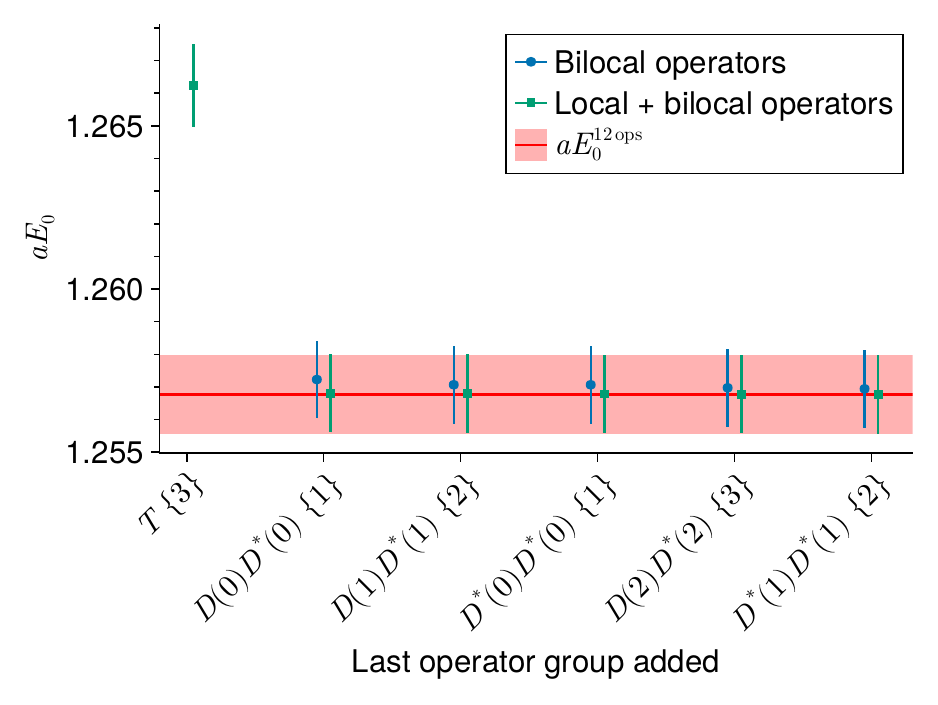}%
  \includegraphics[width=0.5\textwidth]{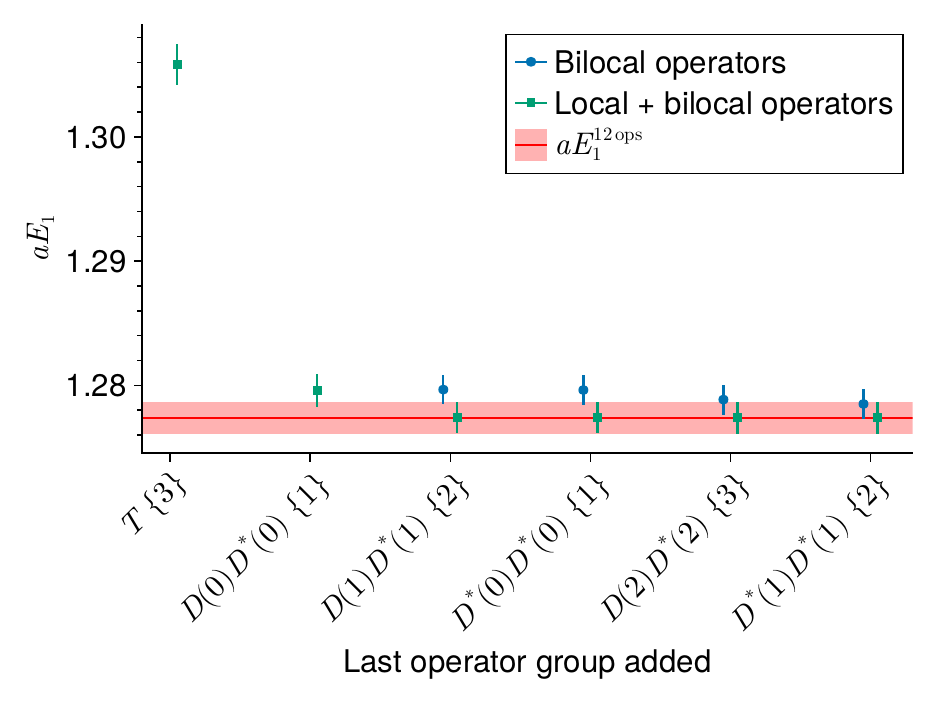}
  
  \includegraphics[width=0.5\textwidth]{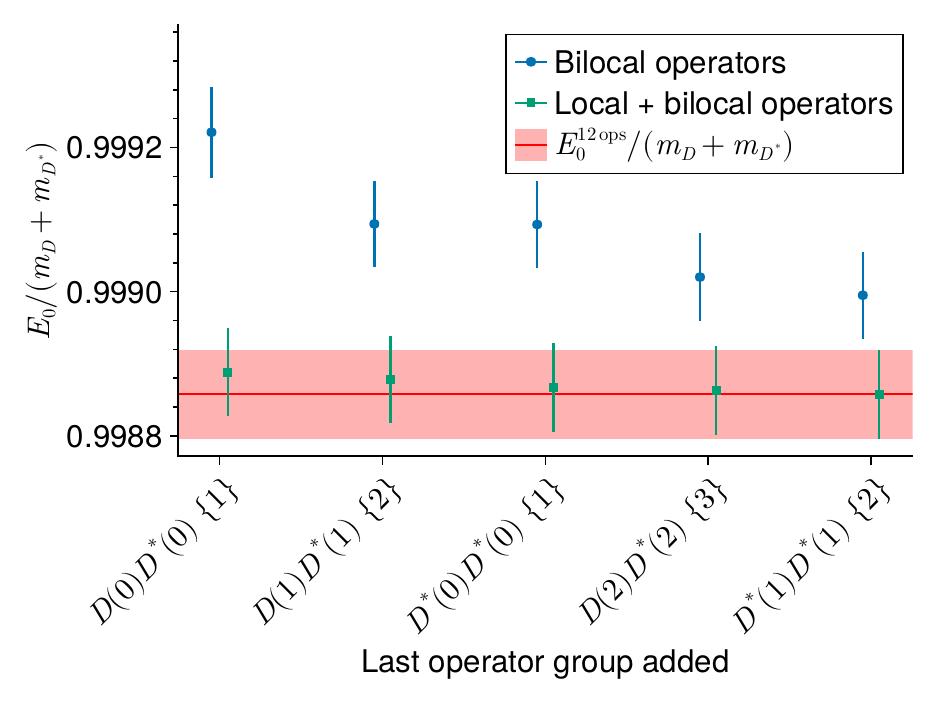}%
  \includegraphics[width=0.5\textwidth]{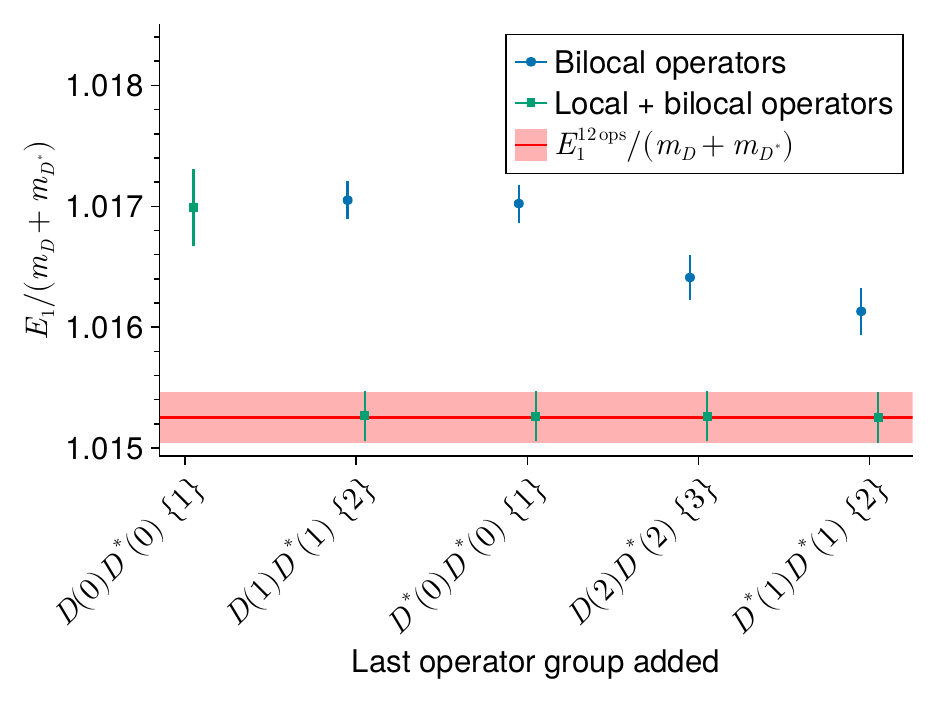}
  \caption{\label{fig:operator_importance}Finite-volume ground state (top left) and first excited state (top right) energy estimates from using different operator bases with and without local tetraquark operators. The x axis denotes the last operator group added to the basis for computing $E_n$; groups to the left were already present (except $T\;\{3\}$ when only bilocal operators are used). The red bands show the results from using all 12 operators. The lower panels show the same results, but the energy estimates are normalized with the threshold energy using correlated ratios. We removed the $T\;\{3\}$ energy estimates from the lower plots due to their large values compared to the others. All energies were computed on the N202 gauge ensemble.}
\end{figure*}

The resulting energy estimates for the different operator bases on the N202 ensemble are shown in the top panels of Fig.~\ref{fig:operator_importance}.
The plot for the ground state energy (top left panel) shows that the inclusion of local tetraquark operators does not change the estimate of this level significantly. The result from just using one bilocal scattering operator already agrees well with the best estimate from using all 12 operators.

For the first excited state (top right panel), we see that the energy estimates from only using bilocal operators decrease in a steplike manner when adding operator groups. On the other hand, when including the local tetraquark operators, the energy estimates have already converged when using the first three operator groups. When using all bilocal operators, the resulting energy estimate is still roughly 1~$\sigma$ away from the 12-operators result. This indicates that the bilocal scattering operators on their own do not capture the correct spectrum well. We also see that the local tetraquark operators on their own do not capture the spectrum correctly: for both the ground and the first excited state, the $T\;\{3\}$ result is far away from the 12-operators result.

In a Lüscher analysis, both the two-particle energy levels and the single-particle masses are important as both enter the calculation. Consequently, a large correlation between the two can result in a significant error cancelation. Therefore, we also analyzed the correlated ratio between the interacting energies and the threshold energy $m_D + m_{D^*}$. The results are displayed in the lower panels of Fig.~\ref{fig:operator_importance}. Compared to the upper panels, we see a large error cancelation, which results in a more significant difference between the result from the bilocal and the full operator basis. For both the ground and the first excited state, we see a steplike decrease in the energy estimates from the bilocal bases, and they do not reach the 12-operators result. On the other hand, the bases with local operators result in energy estimates that stabilize when only including a few operator groups. Consequently, the best estimates from the bilocal and the full operator bases differ significantly. For the ground state, the difference is roughly 2~$\sigma$ and for the first excited state it is approximately 3~$\sigma$. Presumably, adding more $D^{(*)}D^*$ operators on higher momentum shells to the bilocal operator basis would decrease the resulting energies. However, the bilocal operators become increasingly expensive on higher momentum shells due to the larger number of momentum combinations.

This type of operator-importance analysis provides a useful way to test the reliability of the extracted spectrum. Even without comparing the energy estimates from the bilocal and the full basis, one can infer that the first excited state is not reliably determined using only our set of bilocal operators.

To investigate the individual influence of the two types of tetraquark operators on the spectrum, we repeated the operator-importance analysis while omitting either the diquark-antidiquark operator or the local $D^{(*)}D^*$ operators (results not shown). If fewer operator types are sufficient, then the construction of correlation functions could be simplified. For this analysis, we again investigated the ratios of the lowest two energy level estimates to the threshold energy on the N202 ensemble. We found that excluding the local diquark-antidiquark operator from the operator basis yields energy estimates that are virtually identical to those obtained when all three tetraquark operators are included. In contrast, omitting the two local meson-meson operators results in a steplike decrease in the first excited state as more bilocal operators are added; when all nine bilocal operators are used, the remaining difference is about $0.4\;\sigma$. For the ground state, this combination of all bilocal operators and the local diquark-antidiquark operator also produces a negligible difference relative to the full operator set. Consequently, using bilocal meson-meson operators in combination with either the two local meson-meson operators or the diquark-antidiquark operator is sufficient to extract the lowest two energy levels accurately. However, using the local meson-meson operators is more stable with respect to the number of bilocal operators.

\begin{figure*}
  \includegraphics[width=\textwidth]{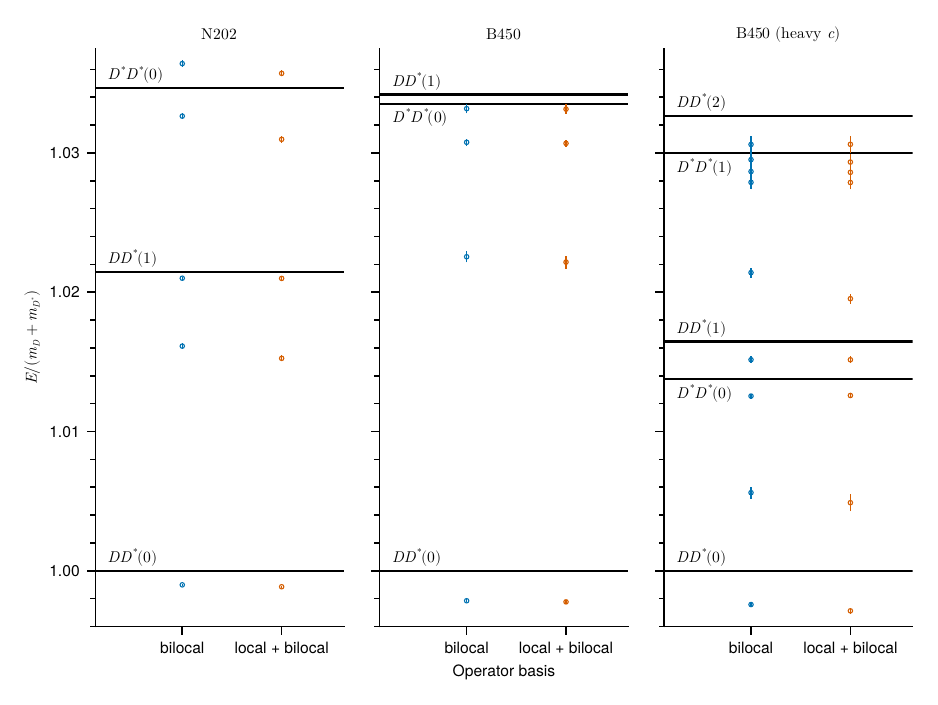}
  \caption{\label{fig:spectrum}Estimate of the finite-volume $cc\overline{u}\overline{d}$, isospin-0 spectrum in the $T_1^+$ irrep for the operator basis consisting only of bilocal scattering operators and for the full operator basis including also local tetraquark operators. The energy estimates are normalized with the threshold energy using correlated ratios. The spectrum is presented for the N202 (left) and the B450 (middle) gauge ensembles. In addition, energy estimates on B450 computed with a heavier-than-physical charm quark are displayed (right). The horizontal lines denote the noninteracting $D^{(*)}D^*$ energies (with the integer back-to-back momentum squared in the parentheses).}
\end{figure*}

Utilizing the basis consisting of all bilocal operators and that of all local and bilocal operators, we computed the estimate of the low-lying finite-volume spectrum on both ensembles. The results are displayed in Fig.~\ref{fig:spectrum}.
For the N202, we see the shifts in the energy estimates for the ground and first excited state that we have already discussed. Besides those two energy levels, there are also significant shifts in higher excited states. Especially the state below the $D^*D^*(0)$ threshold, whose estimate shifts by roughly 5~$\sigma$ upon inclusion of local operators. The only level that is unaffected by the local operators is the state below the $DD^*(1)$ noninteracting level.

On the B450 ensemble, which has a coarser lattice spacing and a smaller physical volume, the inclusion of local operators has a smaller impact. The ground state and the first excited state energy estimates shift by roughly 0.5~$\sigma$. The two levels close to the $D^*D^*(0)$ threshold remain mostly unaffected. The smaller effect of the local operators on B450 could be due to its smaller physical volume, which leads to more widely spaced finite-volume energy levels. Consequently, excited states are more suppressed in the effective energy.

In Fig.~\ref{fig:spectrum} we also show energy estimates on the B450 ensemble that we computed with the heavier than physical charm quark (cf. Chapter~\ref{sec:lattice_setup}).  We include them because the $T_{bb}$ tetraquark, which is deeply bound~\cite{Leskovec:2019ioa, Alexandrou:2024iwi}, arises from substituting the two charm quarks with bottom quarks. Local tetraquark operators were found to be essential for correctly determining the finite-volume spectrum of the $T_{bb}$, particularly its ground state~\cite{Prelovsek:2025vbr}. Therefore, we expect a stronger binding and a bigger importance of local tetraquark operators with our heavy charm quark. This is what we see: the ground state energy estimate shifts by roughly 1.5~$\sigma$ and that of the first excited state by 1~$\sigma$ upon including local operators. There is also a significant shift in the state above the $DD^*(1)$ noninteracting level. This shift was already present with the physical charm quark (not displayed in the plot), but it is enhanced with the heavier charm quark.

We conclude that both the bilocal operators and the basis consisting of local and bilocal operators qualitatively produce the same finite-volume spectrum, i.e.\ one does not miss an energy level when not including local tetraquark operators. But depending on the gauge ensemble, the inclusion of local operators can result in significant shifts in the estimates of some levels. Consequently, not including them can lead to a considerable systematic error.

\subsubsection{$s$-wave Lüscher analysis}

To relate our finite-volume spectra to the infinite-volume $s$-wave $DD^*$ scattering phase shift, we utilized Lüscher's quantization condition~\cite{Luscher:1990ux}. We performed a single-channel $s$-wave analysis, which means we assumed a negligible $d$-wave interaction. In this case, we can obtain the $s$-wave scattering phase shift $\delta_0$ directly from the c.m. scattering momentum $q_\m{cm}$ via the generalized zeta function~\cite{Briceno:2014oea}. In the rest frame, this relation is given by
\begin{equation}
  q_\m{cm}\cot{\delta_0(q_\m{cm})} = \frac{2}{\sqrt{\pi}L} Z_{00}\left(1, \frac{q_\m{cm}L}{2\pi}\right),
\end{equation}
where $q_\m{cm}$ is given in terms of the Mandelstam $s$ via
\begin{equation} \label{eq:luscher_sqrt_s}
	\sqrt{s} = \sqrt{q_\m{cm}^2 + m_D^2} + \sqrt{q_\m{cm}^2 + m_{D^*}^2}.
\end{equation}
Our finite-volume spectra were obtained in the rest frame, so in our case $s = E^2$.

We measured a deviation from the continuum dispersion relation for the $D$ and the $D^*$ meson on the two ensembles that we used. Therefore, we employed a modified dispersion relation given by $\lambda \vb*{p}^2 + m^2$, with a parameter $\lambda \leq 1$. This modification can be derived from Symanzik effective field theory in the limit of small momenta, and it has already been used, e.g. in~\cite{Shi:2025ogt}. It results in the following equation for $q_\m{cm}$:
\begin{equation}
  \sqrt{s} = \sqrt{\lambda_{D}q_\m{cm}^2 + m_D^2} 
    + \sqrt{\lambda_{D^*} q_\m{cm}^2 + m_{D^*}^2}.
\end{equation}
The solution is
\begin{align}
\begin{split}
  q_\m{cm}^2 = &\frac{s}{(\lambda_D - \lambda_{D^*})^2}
    \Bigg[\lambda_D + \lambda_{D^*} + (\lambda_D - \lambda_{D^*})\frac{m_{D^*}^2 - m_D^2}{s} \\
    & \hspace{0.3cm}-2\sqrt{(\lambda_D - \lambda_{D^*})\frac{\lambda_D m_{D^*}^2 - \lambda_{D^*} m_D^2}{s} + \lambda_D\lambda_{D^*}} \,
  \Bigg],
\end{split}
\end{align}
which has the correct limit for $\lambda_D, \lambda_{D^*} \rightarrow 1$.

We obtained the two correction factors $\lambda_D$ and $\lambda_{D^*}$ from a constant fit to $\left(E_{D^{(*)}}(\vb*{p})^2 - m_{D^{(*)}}^2\right)/\vb*{p}^2$ where $E_D(\vb*{p})$ and $E_{D^*}(\vb*{p})$ are the $D$ and $D^*$ energies for momentum $\vb*{p}$. We choose the fit range $(\frac{L}{2\pi})^2\,\vb*{p}^2 = 1, 2, 3$. The resulting values are listed in Table~\ref{tab:dispersion_rel_factors}.
\begin{table}
  \caption{\label{tab:dispersion_rel_factors}Correction factors for the modified dispersion relation.}
  \begin{ruledtabular}
  \begin{tabular}{lcc}
    Ensemble & $\lambda_D$ & $\lambda_{D^*}$ \\
    \hline
    B450 & 0.928(7)  & 0.91(1) \\
    N202 & 0.968(4) & 0.965(6) \\
  \end{tabular}
  \end{ruledtabular}
\end{table}

For single-channel $DD^*$ scattering, Lüscher's quantization condition is valid between the $u$-channel cut coming from the one pion exchange (at $q_\m{cm}^2 \approx - m_\pi^2/4$) and the $D^*D^*(0)$ threshold. Therefore, we only used the lowest two energy levels on both the B450 and the N202 ensemble for our analysis. The third level on N202 could, in principle, also be used, but it is strongly influenced by the $d$ wave.

\begin{figure}
  \includegraphics[width=0.5\textwidth]{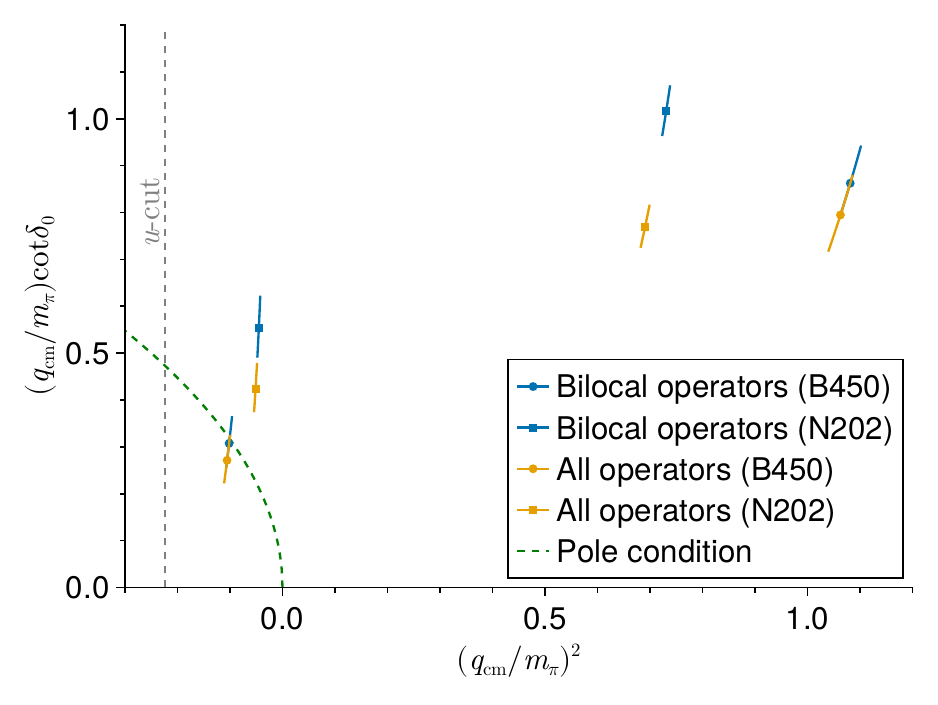}
  \caption{\label{fig:s-wave_analysis}Estimates of the $s$-wave $DD^*$ scattering phase shift in terms of $q_\m{cm}\cot{\delta_0(q_\m{cm})}$ for the full and the bilocal operator basis and for both gauge ensembles. The c.m. scattering momenta $q_\m{cm}$ are normalized with the pion mass $m_\pi$. The green dashed line displays the pole condition $q_\m{cm}\cot{\delta_0(q_\m{cm})} = \pm \sqrt{-q_\m{cm}^2}$ and the vertical gray dashed line shows the start of the $u$-channel cut.}
\end{figure}

The resulting $q_\m{cm}\cot{\delta_0}$ are displayed in Fig.~\ref{fig:s-wave_analysis}.
As expected, the shifts in the estimates of the finite-volume energies upon inclusion of local operators also appear in the estimate of the phase shift. We see that the improved spectrum results in a better agreement between N202 and B450. The analysis without local operators would have suggested a large discretization effect, and this is reduced now.

Our results indicate an intersection of $q_\m{cm}\cot{\delta_0}$ at positive $\sqrt{-q_\m{cm}^2}$, which corresponds to a virtual bound state. However, the presence of the nearby left-hand cut ($u$ cut) may distort this~\cite{Du:2023hlu, Prelovsek:2025vbr, Hansen:2024ffk, Dawid:2024dgy, Dawid:2025wsn}. We plan to investigate this in future work.

\section{Conclusion} \label{sec:conclusion}

We have presented a position-space sampling method for local multiquark operators within the distillation framework that avoids a strong cost scaling of the contractions with the physical volume. It is an unbiased stochastic estimator for correlators that uses randomly displaced sparse grids to compute the momentum projection. It allows combining local multiquark operators and bilocal scattering operators in a variational analysis.

We investigated the variance of this estimator as a function of the point separation in the sparse grids. We did this for single-meson and -baryon operators but also for local tetraquark operators relevant for the $T_{cc}$ tetraquark. For these operators, we can use a large point separation while still getting a statistical error that is dominated by the Monte Carlo error. This makes local multiquark operators affordable in large physical volumes.

For the $T_{cc}$, we analyzed the importance of local tetraquark operators for computing the finite-volume spectrum. To that end, we investigated the convergence of the energy-level estimates upon enlarging a basis consisting of purely bilocal operators and one that also includes local operators. We find qualitative agreement between the spectra obtained from the two bases. However, we observe a fast convergence when using the mixed basis, whereas with the bilocal basis, the energy estimates converge more slowly. Depending on the gauge ensemble, the estimates of several energy levels shift significantly upon inclusion of local operators. Finally, we performed a single-channel $s$-wave Lüscher analysis for the spectra from the bilocal and the full operator basis to compare the resulting $DD^*$ phase shifts.

We conclude that not including local tetraquark operators in an analysis of the $T_{cc}$ can lead to significant systematic errors. Future work will extend this analysis also to the moving frame and will include multiple physical volumes and lattice spacings.


\begin{acknowledgments}
AS is grateful to God for support and useful input throughout this project. We thank Renwick J. Hudspith for code development, and M. Padmanath and Fernando Romero-López for crosschecks.  We are also grateful to our colleagues within the CLS initiative for sharing ensembles. Calculations for this project used resources on the supercomputers JURECA~\cite{krause2018jureca} and JU\-WELS~\cite{krause2019juwels} at Jülich Supercomputing Centre (JSC). The raw distillation data were computed using QDP++~\cite{Edwards:2004sx}, PRIMME~\cite{PRIMME}, and the deflated SAP+GCR solver from openQCD~\cite{openQCD}. Contractions were performed using TensorOperations.jl~\cite{TensorOperations.jl} and ITensors.jl~\cite{ITensor}, the Monte Carlo analysis was done using ADerrors.jl~\cite{ADerrors.jl}, and the plots were prepared with Makie.jl~\cite{Danisch2021}. The tensor network diagrams were generated using TikZ-network~\cite{hackl2018tikznetworkmanual}. AS's research is funded by the Deutsche Forschungsgemeinschaft (DFG, German Research Foundation) - Projektnummer 417533893/GRK2575 ``Rethinking Quantum Field Theory''.
\end{acknowledgments}

\bibliography{References}

\end{document}